\documentclass[aps,prl,twocolumn,showpacs,superscriptaddress,groupedaddress]{revtex4}  % for review and submission
\usepackage{graphicx}  % needed for figures
\usepackage{dcolumn}   % needed for some tables
\usepackage{bm}        % for maths
\usepackage{amssymb}   % for maths
\usepackage{amsmath}   % ditto, for maths
\usepackage{hyperref}
\hypersetup{
     colorlinks=true,
     linkcolor=blue,
     citecolor=blue,      
     urlcolor=blue,
     pdftitle={Calcium peroxide from ambient to high pressures},
}

\begin{document}

\title{Calcium peroxide from ambient to high pressure} 

\author{Joseph R.~Nelson} \email{jn336@cam.ac.uk}

\author{Richard J.~Needs} \affiliation{Theory of Condensed Matter Group,
  Cavendish Laboratory, J.~J.~Thomson Avenue, Cambridge CB3 0HE,
  United Kingdom}

\author{Chris J.~Pickard} \affiliation{Department of Physics and Astronomy,
  University College London, Gower Street, London WC1E 6BT,
  United Kingdom} \vskip 0.25cm

\date{\today}

\begin{abstract}
Structures of calcium peroxide (CaO$_2$) are investigated in the pressure range 0-200 GPa using the \textit{ab initio} random structure searching (AIRSS) method and density functional theory (DFT) calculations. At 0 GPa, there are several CaO$_2$ structures very close in enthalpy, with the ground-state structure dependent on the choice of exchange-correlation functional. Further stable structures for CaO$_2$ with $C2/c$, $I4/mcm$ and $P2_1/c$ symmetries emerge at pressures below 40 GPa. These phases are thermodynamically stable against decomposition into CaO and O$_2$. The stability of CaO$_2$ with respect to decomposition increases with pressure, with peak stability occurring at the CaO B1-B2 phase transition at 65 GPa. Phonon calculations using the quasiharmonic approximation show that CaO$_2$ is a stable oxide of calcium at mantle temperatures and pressures, highlighting a possible role for CaO$_2$ in planetary geochemistry. We sketch the phase diagram for CaO$_2$, and find at least five new stable phases in the pressure/temperature ranges $0\leq P\leq 60$ GPa, $0\leq T\leq 600$ K, including two new candidates for the zero-pressure ground state structure.
\end{abstract}

\maketitle

\section{Introduction} \label{intro}
The typical oxide formed by calcium metal is calcium oxide, CaO, having Ca and O in +2 and -2 oxidation states respectively. Calcium and oxygen can also combine to form calcium peroxide, CaO$_2$, a compound which enjoys a variety of uses in industry and agriculture. Calcium peroxide is used as a source of chemically bound but easily evolved oxygen in fertilisers, for oxygenation and disinfection of water, and in soil remediation \cite{QZZZC2013,MSM2010}.

At ambient pressure bulk calcium peroxide decomposes at a temperature of about 620 K \cite{MSM2010,foot2}. Early X-ray diffraction (XRD) experiments assigned a tetragonal `calcium carbide' structure of space group $I4/mmm$ to CaO$_2$ \cite{CN1956}. This same structure was already known to be formed by heavier alkaline earth metal peroxides \cite{KSC1999,V1993}. Recently, Zhao \textit{et al.} \cite{Z2013} used an adaptive genetic algorithm and density functional theory (DFT) calculations to search for structures of CaO$_2$, finding a new orthorhombic ground state structure of $Pna2_1$ symmetry, which is calculated to be close to thermodynamic stability at zero pressure and temperature. The simulated XRD pattern from this structure is in good agreement with the available experimental data \cite{Z2013}. Thermodynamic stability in this case means stability against decomposition via
\begin{eqnarray}
\label{01}
\mbox{CaO}_2 \longrightarrow \mbox{CaO} + \frac{1}{2}\mbox{O}_2.
\end{eqnarray}

We are interested in the stabilities of structures of CaO$_2$ from zero pressure up to 200 GPa, and temperatures up to 1000 K.  Calcium and oxygen have high abundances in the Earth's crust and mantle and, because they also have high cosmic abundances, stable compounds formed from these elements at high pressures are key (exo)-planetary building blocks. Understanding the structures of such compounds allows insight into the composition of planetary interiors, including exoplanets. To date, almost no work has been performed investigating CaO$_2$ as a stable oxide of calcium at high pressures. Some previous work has explored the effect of low pressures ($<$10 GPa) on the bond lengths and lattice parameters of $I4/mmm$-CaO$_2$ \cite{KSC1999}. We therefore employ DFT calculations to explore the behaviour of CaO$_2$ at pressures in the GPa range. DFT calculations provide an excellent avenue for investigating materials properties under pressure, both at pressures accessible to diamond anvil cells \cite{Ninet_ionic_ammonia_2014} and at terapascal pressures \cite{MPN2012,Pickard_terapascal_water_2013}. To explore the stability of CaO$_2$, we search for new crystal structures of this compound at a variety of pressures in the range 0-200 GPa.

\section{Methods} \label{methods}
Density functional theory calculations are performed using the \textsc{castep} plane-wave pseudopotential code \cite{CS2005}. Ultrasoft pseudopotentials \cite{V1990} generated with the \textsc{castep} code are used for both calcium and oxygen, with core states $1s^22s^22p^6$ and $1s^2$, respectively. We use the Perdew-Burke-Ernzerhof (PBE) \cite{PBE1996} form of the exchange-correlation functional with a plane-wave basis cutoff of 800 eV. A $k$-point sampling density of $2\pi\times0.03$ \AA$^{-1}$ is used for our CaO and CaO$_2$ phases. For our oxygen phases, we use a denser $k$-point sampling of $2\pi\times0.02$ \AA$^{-1}$. Bulk modulii are calculated by fitting static lattice pressure-volume data to the third-order Birch-Murnaghan equation of state.

The electronic density of states is calculated using the \textsc{optados} code \cite{optados1,optados2,optados3}. Calculations of phonon frequencies are performed with the \textsc{castep} code and a finite-displacement supercell method, using the quasiharmonic approximation \cite{KW2003,Carrier}.

To search for new phases of CaO$_2$, we use the \textit{ab initio} random structure searching (AIRSS) technique \cite{PN2011}. AIRSS proceeds by generating random starting structures containing a given number of formula units. Of these, structures which have lattice parameters giving reasonable bond lengths and cell volumes at a particular pressure are then relaxed to an enthalpy minimum, and the lowest enthalpy structures are selected for refinement.  AIRSS has proved to be a very powerful tool in predicting new structures, several of which have subsequently been found experimentally. AIRSS searches have for example uncovered high pressure phases of silane \cite{PN2006}, and correctly predicted high pressure metallic phases in aluminium hydrides \cite{PN2007}.

In this study, we perform structure searching at pressures of 0, 10, 20, 50, 100, 150 and 200 GPa. The bulk (about 60\%) of our searches use cells with 2 or 4 formula units of CaO$_2$, and we have also performed searches with cells containing 1, 3, 5, 6 and 8 formula units. Not all combinations of pressures and formula unit numbers are searched. In total, we relax over 25,000 structures in our CaO$_2$ searches. We supplement our AIRSS searches by calculating the enthalpies of five known alkaline earth metal peroxide structures taken from the Inorganic Crystal Structure Database (ICSD) \cite{ICSD}, and other authors \cite{Z2013,KC1998,ZO2013}. The relevant alkaline earth metal is replaced with calcium where appropriate.

\begin{figure*}
\centering
  \includegraphics[scale=0.64,clip]{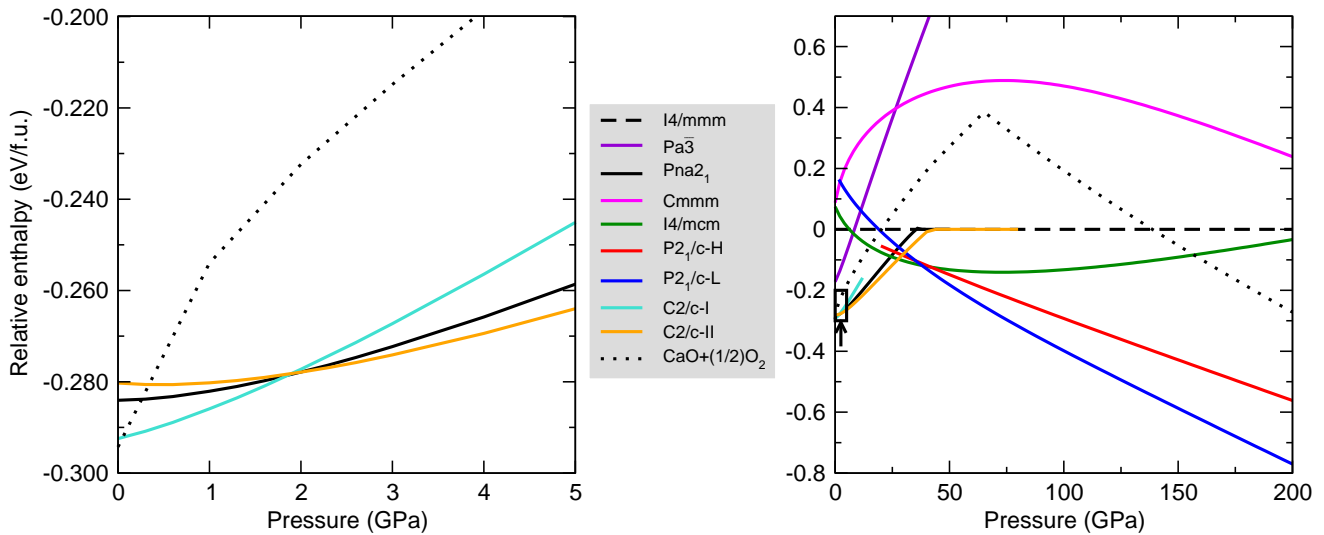}
  \caption{\label{fig:Enthalpies} Static lattice enthalpies, in eV per unit of CaO$_2$, of calcium peroxide phases in the pressure range 0-5 GPa (\textit{left}) and 0-200 GPa (\textit{right}). Enthalpies are given relative to the $I4/mmm$ phase of CaO$_2$. $C2/c$-I, $C2/c$-II, $P2_1/c$-$H$ and $P2_1/c$-$L$ are new structures of CaO$_2$ found using AIRSS. The left-hand plot highlights the enthalpy differences between the $Pna2_1$ and two $C2/c$ phases at low pressures; $C2/c$-I is the lowest-enthalpy structure at 0 GPa from our searches. The arrow and box in the right-hand figure indicate the scope of the left-hand figure. Structures of CaO$_2$ below the dotted line are thermodynamically stable against decomposition into CaO and O$_2$.}
\end{figure*}

\subsection{CaO}
CaO undergoes a transition from the rocksalt ($Fm\bar{3}m$) to the CsCl ($Pm\bar{3}m$) structure (the B1-B2 transition) around 60-65 GPa. Diamond-anvil-cell experiments indicate a transition pressure of 60$\pm$2 GPa at room temperature \cite{JAMB1979}, while DFT calculations give a transition pressure of 65-66 GPa \cite{C2003,ZK2009}. We find a pressure of 65 GPa in the present study, and we therefore use the $Fm\bar{3}m$ rocksalt structure below 65 GPa and the CsCl structure at higher pressures. The bulk modulus of $Fm\bar{3}m$-CaO has been measured to be 104.9 GPa \cite{soga}, while our static-lattice DFT calculations give a value of 108.5 GPa. To exclude the possibility that CaO might have a different, more stable structure (other than $Fm\bar{3}m$ or $Pm\bar{3}m$) over the pressure range 0-200 GPa, we also perform AIRSS on CaO at pressures of 50, 100 and 200 GPa. We do not find any new low-enthalpy structures for CaO at these pressures.

\subsection{Solid oxygen}
Structure searching over the pressure range 0-200 GPa has already been performed for solid oxygen \cite{SMK2012, M2007}, and we use the lowest enthalpy structures found therein. We also examine the enthalpies of the experimentally-determined $\alpha$ and $\delta$ oxygen phases at low pressures \cite{NeatonandAshcroft}. Choosing the lowest-enthalpy oxygen structure at each pressure, we find that $\delta$-O$_2$ is stable between 0 and 1.2 GPa, after which an insulating phase with space group $Cmcm$ \cite{NeatonandAshcroft} is stable up to 41 GPa. A phase of symmetry $C2/m$ \cite{M2007} then becomes stable, remaining so up to 200 GPa. Our spin-polarised calculations show no discernable difference in enthalpy between a $\delta$-O$_2$ phase with antiferromagnetic spin ordering, and the experimentally-determined ferromagnetic spin-ordering for $\delta$-O$_2$ \cite{Goncharenko}.

These results are not in accord with low-temperature experiments on solid oxygen, which predict $\alpha$-O$_2$ to be stable between 0 and about 5 GPa, $\delta$-O$_2$ to be stable between about 5 and 10 GPa, followed by the `$\epsilon$-O$_2$' phase between 10 and 96 GPa, with a further isostructural phase transition around 100 GPa \cite{LWMDL2006, F2006, AKHH1995}. Our DFT calculations do not yield $\alpha$-O$_2$; optimising its structure at low pressures simply gives the $\delta$-O$_2$ structure. The aforementioned $C2/m$ oxygen phase is however very similar in structure to $\epsilon$-O$_2$ (which is also of $C2/m$ symmetry), and is within 50 meV/f.u. of that phase in enthalpy over the pressure range 0-200 GPa. Any higher enthalpy structure for oxygen over the pressure range being explored here would only increase the calculated stability of CaO$_2$, so we proceed with the lowest enthalpy DFT phases for oxygen.

\section{\label{results}Results}
\subsection{\label{strucs}Structure searching and static lattice results}
Our structure searching is carried out by minimising the enthalpy at a given pressure within the static-lattice approximation. We discuss these results first before presenting our calculations of the Gibbs free energy.

Fig.~\ref{fig:Enthalpies} shows the enthalpy-pressure curves for nine phases of CaO$_2$. The first five of these, labelled with their space group symmetries $I4/mmm$, $Pa\overline{3}$, $Pna2_1$, $Cmmm$ and $I4/mcm$ in Fig.~\ref{fig:Enthalpies}, are known alkaline earth metal peroxide structures as mentioned in the `Methods' section. The other four, with space group symmetries $C2/c$ and $P2_1/c$, are new CaO$_2$ structures. These are the lowest-enthalpy phases that turned up during our AIRSS searches. The dotted line in Fig.~\ref{fig:Enthalpies} shows the enthalpy of CaO+$\frac{1}{2}$O$_2$, calculated using the lowest-enthalpy phases of CaO and O$_2$ at each pressure. Any CaO$_2$ phase below this dotted line is stable against decomposition in the manner of Eq.~(\ref{01}).

We find the following sequence of phase transitions for CaO$_2$ by considering the lowest enthalpy structure at each pressure:
\begin{eqnarray*}
&C2/c\mbox{-I} \xrightarrow{\mbox{1.8 GPa}} Pna2_1 \xrightarrow{\mbox{2.0 GPa}} C2/c\mbox{-II} \\
& \xrightarrow{\mbox{27.6 GPa}} I4/mcm \xrightarrow{\mbox{37.9 GPa}} P2_1/c\mbox{-}L, 
\end{eqnarray*}
with the transition pressures between each structure as indicated by the arrows. The $P2_1/c$-$L$ phase is then predicted to be stable up to at least 200 GPa.

At 0 GPa, our structure searching reveals a number of CaO$_2$ phases which are very close (within 10 meV/f.u.) in enthalpy. Within the present approximations (DFT with PBE exchange-correlation), a phase of $C2/c$ symmetry (`$C2/c$-I') is lowest in enthalpy at 0 GPa. This phase is 8.4 meV/f.u.~lower in enthalpy than the proposed $Pna2_1$-symmetry ground state structure of Zhao \textit{et al.} \cite{Z2013}, which we note also turned up in our AIRSS searches. Our searches also uncover a second $C2/c$-symmetry phase (`$C2/c$-II') which is slightly higher in enthalpy than $Pna2_1$ at 0 GPa. The enthalpies of these three phases at low pressures are shown in more detail in the left-hand panel of Fig.~\ref{fig:Enthalpies}.

We find that the calculated XRD patterns for the $C2/c$-I, $C2/c$-II and $Pna2_1$ structures share many similar features with available experimental data \cite{KC1998}, although this XRD data is best fitted by the $Pna2_1$ structure. The enthalpy differences between the $C2/c$-I, -II and $Pna2_1$ phases at 0 GPa are dependent on the choice of exchange-correlation functional. In Table \ref{tab:tablee}, we show the enthalpies of the $Pna2_1$ and $C2/c$-II phases relative to the $C2/c$-I phase using a variety of different functionals. We find differences in calculated equilibrium volumes as well: for $C2/c$-I at 0 GPa, the LDA gives a volume of 35.8 \AA$^3$/f.u., while the PBE functional gives 39.1 \AA$^3$/f.u. The LDA typically overbinds in DFT calculations, with PBE instead underbinding, so these two volumes likely bracket the true volume of the $C2/c$-I phase at 0 GPa. Given that $dP/dV$ is about -2.3 GPa/\AA$^3$ for this phase at 0 GPa, this volume difference (due to functional choice) corresponds to a pressure uncertainty in the neighbourhood of $\pm3$ GPa. In light of this uncertainty, $C2/c$-I,$C2/c$-II and $Pna2_1$ are all reasonable candidates for the structure of CaO$_2$ at 0 GPa.

\begin{table}[h]
\caption{\label{tab:tablee}Enthalpies of the $C2/c$-II and $Pna2_1$ phases of CaO$_2$ at 0 GPa, in meV per unit of CaO$_2$, relative to the $C2/c$-I phase using different exchange-correlation functionals.}
\begin{tabular}{lrrrrr} \hline
Functional  & LDA    & PBE   & PBESOL & PW91  & WC \\ \hline
$C2/c$-I    & 0.0   & 0.0  & 0.0   & 0.0  & 0.0 \\
$Pna2_1$    & 0.1   & 8.4  & 2.9   & 8.7  & 3.5 \\
$C2/c$-II   & -16.3 & 12.2 & -4.1  & 12.0 & -2.7 \\ \hline
\end{tabular}

\end{table}

\begin{figure}
\centering
  \includegraphics[scale=0.23]{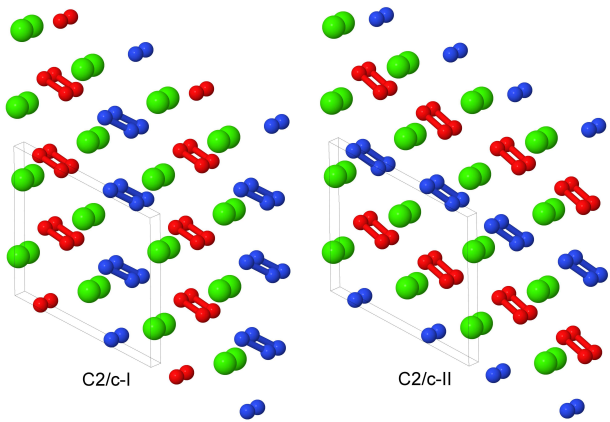}
  \caption{\label{fig:cIandII} 2x2x2 slabs of the $C2/c$-I (\textit{left}) and $C2/c$-II (\textit{right}) structures, viewed almost down the $b$-axis. Both structures are very similar, with Ca atoms forming an almost-hexagonal motif when viewed from this angle. Green atoms correspond to Ca atoms, while red and blue correspond to O atoms. All O atoms in peroxide ions with parallel O-O axes are given the same colour.}
\end{figure}

The $C2/c$-I, -II and $Pna2_1$ phases exhibit very similar structures. Looking down the $b$-axis of each phase, calcium atoms form a nearly-hexagonal motif, and we note that the monoclinic angle $\beta$ is close to 120$^{\circ}$ for $C2/c$-I and $C2/c$-II. For the $C2/c$-I and -II phases, the peroxide ion axes are almost coplanar, as can be seen in Fig.~\ref{fig:cIandII}. In $C2/c$-II, the axes of peroxide ions in the same plane are parallel, while in $C2/c$-I, they alternate in orientation in the same plane. We use red and blue colouring in Fig.~\ref{fig:cIandII} to show peroxide ions with parallel axes.

AIRSS searches also reveal a second phase for CaO$_2$ with $P2_1/c$ symmetry (`$P2_1/c$-$H$'). As can be seen in the right-hand panel of Fig.~\ref{fig:Enthalpies}, our static-lattice calculations show that this is not a stable phase for CaO$_2$ over the pressure range 0-200 GPa. However around 38 GPa, the enthalpy of this phase lies within 10 meV/f.u. of the $I4/mcm$ and $P2_1/c$-$L$ phases, opening up the possibility this phase could become stable once we take temperature into account. We consider this further in a later section (`Lattice dynamics'). The $I4/mcm$ phase reported here is also predicted for MgO$_2$ above 53 GPa \cite{ZO2013}.

We provide lattice parameters and bulk modulii for the $C2/c$-I, $C2/c$-II, $I4/mcm$, $P2_1/c$-$H$ and $P2_1/c$-$L$ phases at 0, 20, 30 and 50 GPa in Table \ref{tab:table1}.

\begin{table*}
\caption{\label{tab:table1}Structures of the $C2/c$-I, $C2/c$-II, $I4/mcm$, $P2_1/c$-$H$ and $P2_1/c$-$L$ phases of CaO$_2$.}
\begin{tabular}{llllllccccc} \hline
 Pressure& &\multicolumn{3}{c}{Lattice parameters}& & \multicolumn{3}{c}{Atomic coordinates} & Wyckoff &  \\
 (GPa)&Space group& &(\AA, deg.)& &Atom&$x$&$y$&$z$& site & $B_0$ (GPa) \\ \hline
 0   &$C2/c$ (\#15)$^{a}$ (I) &$a$=7.041    &$b$=3.685    &$c$=6.820    &Ca&0.0000&0.6399&0.2500&4e&87 \\
      &      &$\alpha$=90.0&$\beta$=117.8 &$\gamma$=90.0&O &0.2548&0.1404&0.4119&8f& \\
\rule{0pt}{3ex}20&$C2/c$ (\#15)$^{a}$ (II) &$a$=6.829    &$b$=3.403    &$c$=6.407    &Ca&0.0000&0.3413&0.2500&4e&89 \\
      &      &$\alpha$=90.0&$\beta$=118.8 &$\gamma$=90.0&O &0.1661&0.1553&0.0252&8f& \\
\rule{0pt}{3ex}30&$I4/mcm$ (\#140) &$a$=4.521    &$b$=4.521    &$c$=5.745    &Ca&0.0000&0.0000&0.2500&4a&114 \\
      &      &$\alpha$=90.0&$\beta$=90.0&$\gamma$=90.0 &O &0.1143&0.6143&0.0000&8h& \\
\rule{0pt}{3ex}30&$P2_1/c$-$H$ (\#14)$^{b}$ &$a$=6.590    &$b$=4.842    &$c$=3.795    &Ca&0.0611&0.7628&0.2781&4e&93 \\
    &      &$\alpha$=90.0&$\beta$=105.2 &$\gamma$=90.0&O &0.1301&0.2714&0.3087&4e& \\
    &      &             &             &             &O &0.2446&0.4378&0.1012&4e& \\
\rule{0pt}{3ex}50&$P2_1/c$-$L$ (\#14)$^{b}$ &$a$=4.223    &$b$=4.279    &$c$=5.949    &Ca&0.0613&0.5133&0.2539&4e&110 \\
    &      &$\alpha$=90.0&$\beta$=98.7 &$\gamma$=90.0&O &0.0820&0.0235&0.4019&4e& \\
    &      &             &             &             &O &0.1224&0.1182&0.0089&4e& \\ \hline
\multicolumn{10}{l}{$^{a}$ \footnotesize{$C12/c1$ -  International Tables, Volume A: unique axis $b$, cell choice 1.}} \\
\multicolumn{10}{l}{$^{b}$ \footnotesize{$P12_1/a1$ - International Tables, Volume A: unique axis $b$, cell choice 3.}}
\end{tabular}
\end{table*}

\subsection{Bonding and electronic structure}
\begin{figure}[h]
  \includegraphics[scale=0.45,clip]{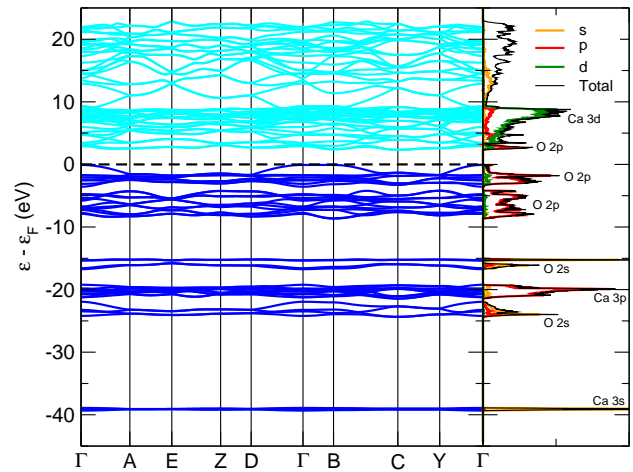}
  \caption{\label{fig:50band} Bandstructure and electronic density of states of the $P2_1/c$-$L$ phase at 50 GPa. The density of states is shown projected onto the $s$, $p$ and $d$ angular momentum channels. We calculate a thermal bandgap of 2.4 eV, and an optical bandgap of 2.5 eV. The Fermi level is shown as a black dashed line.}
\end{figure}

The bandstructure of the $P2_1/c$-$L$ phase at 50 GPa is shown in Fig.~\ref{fig:50band}. The insulating nature of this phase is evident, with a calculated thermal bandgap of 2.4 eV and optical bandgap of 2.5 eV. These will be underestimates of the true bandgap owing to the use of the PBE functional. The lowest set of 4 bands is comprised almost entirely of Ca $3s$ orbitals. Above this, we find a central peak flanked by two smaller sidepeaks in the total density of states. The central peak is built from 12 bands and is largely Ca $3p$ orbitals, while the two sidepeaks contain 4 bands apiece and are dominated by O $2s$ orbitals. Thus, we have a splitting of the O $2s$ orbital energy levels, possibly arising from the covalent bonding present in the peroxide [O-O]$^{-}$ ion. Above this, and just below the HOMO (highest occupied molecular orbital), are 20 bands which arise largely from O $2p$ orbitals. The first orbitals above the Fermi level consist of (unoccupied) O $2p$ orbitals, followed by a dense band of Ca $3d$ orbitals.  An almost identical pattern of bonding and electronic density of states are found in our other low-enthalpy phases.

\begin{figure}[h]
\centering
  \includegraphics[scale=0.45,clip]{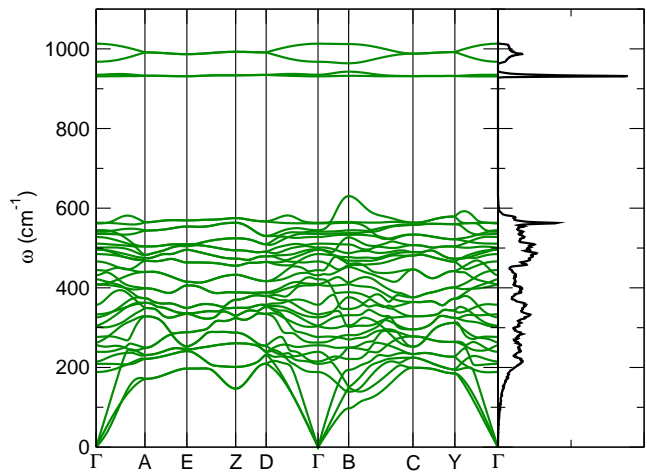}
  \caption{\label{fig:50phonon} Phonon dispersion relations and density of states for $P2_1/c$-$L$ CaO$_2$ at 50 GPa.}
\end{figure}

Fig.~\ref{fig:50phonon} shows the corresponding phonon band structure and density of states for $P2_1/c$-$L$ at 50 GPa. The lack of imaginary phonon frequencies indicates the stability of this particular phase. The distinctly separate high frequency bands around 930 - 1000 cm$^{-1}$ in Fig.~\ref{fig:50phonon} correspond to phonon modes that stretch the O-O covalent bond in the peroxide ions. Two distinct peroxide bond lengths are found in the ions of this structure: 1.44 \AA\ and 1.46 \AA, which splits these higher frequency bands. These peroxide O-O bond lengths are somewhat longer those found in molecular oxygen, which has an O-O bond length of 1.207 \AA\ \cite{KC1998} at ambient conditions, and the bond length in $C2/m$ oxygen at 50 GPa, which we calculate as 1.20 \AA. These longer O-O bond lengths are however typical of crystalline ionic peroxides \cite{KSC1999}.

\subsection{Stability of CaO$_2$}
As can be seen in Fig.~\ref{fig:Enthalpies}, our low-enthalpy phases for CaO$_2$ show remarkable stability against decomposition as pressure increases. We predict CaO$_2$ to be most stable at the B1-B2 CaO phase transition pressure of 65 GPa. There, the decomposition enthalpy (Eq.~(\ref{01})) of CaO$_2$ is +0.64 eV/unit of CaO$_2$ (+62 kJ/mol).

One implication of the stability of CaO$_2$ is that it may be preferentially formed over CaO in an oxygen-rich environment under pressure, through the reverse of Eq.~(\ref{01}). For example, the pressure in the Earth's lower mantle, at a depth of around 1550 km, is about 65 GPa \cite{JAMB1979}, close to our predicted peak stability pressure for CaO$_2$. The formation of MgO$_2$ in this way has also been discussed \cite{ZO2013}, although much higher pressures ($>$116 GPa) are needed before MgO$_2$ is stable against decomposition, whereas CaO$_2$ is stable from around 0 GPa. The mantle temperature in the Earth at 65 GPa is in the neighbourhood of 2500 K, and our phonon calculations show that under these conditions, $\Delta G = +0.54$ eV/f.u. for the reaction of Eq.~(\ref{01}). Hence, CaO$_2$ is a thermodynamically stable oxide at temperatures and pressures encountered in planetary interiors.

Reactions of the form $X + $O$_2 \rightarrow Y$ for species $X$ and $Y$, such as the reverse of Eq.~(\ref{01}), are known as redox buffers. Such reactions are key in determining planetary mantle compositions. In the Earth's mantle, there are a number of such buffers, usually involving the further oxidation of iron and nickel compounds, such as Fe$_3$O$_4$ + $\frac{1}{4}$O$_2$ $\longrightarrow$ $\frac{3}{2}$Fe$_2$O$_3$ and Ni + $\frac{1}{2}$O$_2$ $\longrightarrow$ NiO. The natural formation of CaO$_2$ by further oxidation of CaO in Earth's mantle, while energetically favourable at high pressures and temperatures, would also need to compete against the further oxidation of other such compounds. The average CaO content in the Earth's mantle is about 3\% \cite{soga}, though exoplanet mantles offer a rich variety of alternative compositions.

We highlight the fact that the pressures at which these different CaO$_2$ phases are predicted to become stable are amenable to experimental study in diamond anvil cells. Very few pressure-induced phase transitions for A[B$_2$] compounds are experimentally known \cite{E2010}, although at least one example already occurs among the alkaline earth metal peroxides, in BaO$_2$ \cite{E2010}. Equivalently, it would be interesting to test the reactivity of CaO and O$_2$ under conditions of excess oxygen, as our results suggest that CaO and O$_2$ are reactive at high pressures. The formation of CaO$_2$ may require laser heating in a diamond anvil cell to overcome the likely high potential barriers between phases. High pressure phases of CaO$_2$ could be recoverable at lower pressures, although possibly not at ambient or zero pressure.

\begin{figure}
\centering
  \includegraphics[scale=0.48,clip]{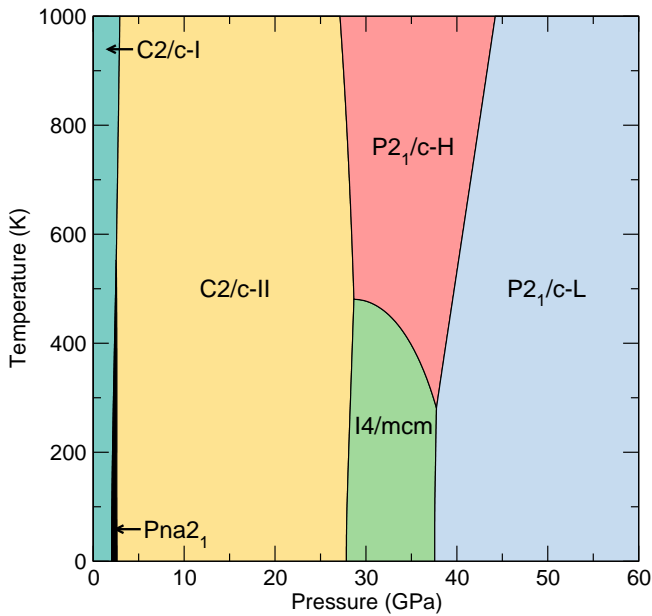}
  \caption{\label{fig:PhaseD} T-P phase diagram of CaO$_2$ as calculated using the quasiharmonic approximation and our lowest-enthalpy structures.}
\end{figure}

\subsection{\label{latticed}Lattice dynamics}
In addition to our static-lattice calculations, we calculate the phonon free energies of our lowest-enthalpy CaO$_2$ phases, namely those with $Pna2_1$, $C2/c$-I/II, $I4/mcm$, $P2_1/c$-$H$ and $P2_1/c$-$L$ symmetries. We encounter no imaginary phonon frequencies over the pressure ranges relevant to these phases, indicating that they are dynamically stable. The relevant thermodynamic potential is now the Gibbs free energy, which includes the phonon pressure. Selecting the lowest Gibbs free energy structure from these six structures gives rise to the T-P phase diagram given in Fig.~\ref{fig:PhaseD}. We note that the upper-left (low-P, high-T) part of the phase diagram is representative only, because at ambient pressures CaO$_2$ decomposes at temperatures around 620 K \cite{MSM2010,foot2}. 

We find that the small enthalpy difference between the $I4/mcm$ and $P2_1/c$-$H$ phases seen in our static-lattice calculations (Fig.~\ref{fig:Enthalpies}) closes with increasing temperature, and we see the emergence of $P2_1/c$-$H$ as a stable phase for CaO$_2$ at 37.7 GPa and for $T>$ 281 K. The free energy difference between the $I4/mcm$ and $P2_1/c$-$H$ phases does remain quite small however, around 20 meV per CaO$_2$ unit at the most over the temperature range 0-1000 K.

At room temperature (300 K), we therefore predict a different sequence of phase transitions than those for our static-lattice calculations. We find that, with PBE exchange-correlation:
\begin{eqnarray*}
&C2/c\mbox{-I} \xrightarrow{\mbox{2.2 GPa}} Pna2_1 \xrightarrow{\mbox{2.6 GPa}} C2/c\mbox{-II} \xrightarrow{\mbox{28.3 GPa}} \\
&I4/mcm \xrightarrow{\mbox{37.4 GPa}} P2_1/c\mbox{-}H \xrightarrow{\mbox{37.9 GPa}} P2_1/c\mbox{-}L, 
\end{eqnarray*}
with the arrows labelled by the predicted transition pressures.

The phase diagram of Fig.~\ref{fig:PhaseD} does not extend all the way to 200 GPa. However, we expect $P2_1/c$-$L$ to continue to be the most stable phase at high pressures. This is because our structure searching results (which use static-lattice enthalpies) show that the next most stable structure for CaO$_2$ over the pressure range 100-200 GPa is at least 45 meV per unit of CaO$_2$ higher in enthalpy. This was not the case at low pressures, where our searches reveal quite a few structures (such as $P2_1/c$-$H$) that are close to becoming stable and may therefore do so at high temperatures.

\section{Conclusions}
Structural changes in CaO$_2$ under pressure have been explored over the pressure range 0-200 GPa at temperatures up to 1000 K. CaO$_2$ remains insulating up to pressures of at least 200 GPa. Structure searching and DFT calculations reveal six stable phases for CaO$_2$ over these pressure and temperature ranges, of which five are reported for the first time in this study. Calculations of the phonon frequencies of these new structures confirms their dynamical stability. The lowest-enthalpy phase of CaO$_2$ at 0 GPa is dependent on the choice of DFT exchange-correlation functional. At pressures above 40 GPa, a phase of $P2_1/c$ symmetry (`$P2_1/c$-$L$') emerges for CaO$_2$ which is predicted to be stable up to 200 GPa, and at mantle pressures and temperatures. CaO$_2$ is a very stable oxide of calcium at high pressures, and may be a constituent of exoplanet mantles. The pressures at which these CaO$_2$ phases become stable are readily attainable in diamond anvil cells.

\section{Acknowledgements}
Calculations were performed using the Darwin Supercomputer of the University of Cambridge High Performance Computing Service (\url{http://www.hpc.cam.ac.uk/}), as well as the ARCHER UK National Supercomputing Service (\url{http://www.archer.ac.uk/}). Financial support was provided by the Engineering and Physical Sciences Research Council (UK). JRN acknowledges the support of the Cambridge Commonwealth Trust.


\begin{thebibliography}{99}
\bibitem{foot2}
This decomposition temperature may be lower for samples of different purities; for example the CRC Handbook \cite{CRC} reports a decomposition temperature of $\approx$470K for CaO$_2$.
\end{thebibliography}
\end{document}